# PDF ARTICLE METADATA HARVESTER

**Leon Andretti Abdillah**

Information Systems, Computer Science Faculty, Bina Darma University,
Jl. A. Yani No.12, Palembang 30264, Indonesia
E-mail: leon.abdillah@yahoo.com

*Abstract*

*Scientific journals are very important in recording the finding from researchers around the world. The recent media to disseminate scientific journals is PDF. On scheme to find the scientific journals over the internet is via metadata. Metadata stores information about article summary. Embedding metadata into PDF of scientific article will grant the consistency of metadata readness. Harvesting the metadata from scientific journal is very interesting field at the moment. This paper will discuss about scientific journal metadata harvesters involving XMP.*

*Abstrak*

*Jurrnal ilmiah sangat penting dalam menyimpan penemuan para peneliti di seluruh dunia. Saat ini media penyimpan artikel ilmiah adalah PDF. Selanjutnya, untuk menemukan jurnal ilmiah di intenet adalah melalui metadata. Metadata menyimpan informasi tentang kesimpulan artikel. Dengan melekatkan metadata pada artikel ilmiah yang berbentuk PDF akan menjamin konsistensi pembacaan metadata. Pengumpulan metadata dari jurnal ilmiah adalah bidang yang sangat menarik deuasa ini. Paper ini akan mendiskusikan pengumpulan metadata pada jurnal ilmiah yang melibatkan XMP.*

*Kata kunci: Artikel jurnal ilmiah, metadata, harvester, XMP*

## INTRODUCTION

Every day, each publisher and/or author(s) compete to publish their new papers, scientific literatures, through the cloud. Scientific literatures published in both manuscript format and available electronically. The basic task is to make these documents searchable and retrievable [1]. In the field of digital library, scientific workers always search a lot of scientific documents at the domain of their researches [2]. They will work for new idea, invention, etc based on the previous research or to deal with the current and future challenges.

The electronic representation of scientific documents may include journals, technical reports, program documentation, laboratory notebooks etc [3]. The most popular of scientific documents is scientific journals article of a particular field or topic.

An article from scientific journal, commonly dominated by word(s) or text(s), but to clarify the discussion then it could be added by several forms such as black-and-white line(s), chart(s), diagram(s), equation(s), formula(s), graphic(s), illustration(s), photograph(s), picture(s), table(s), etc. Scientific journals have a formal structure that has to be understood by all those who read it (authors, readers and editors) in order to be useful [4], especially for editors who will interact directly to the article. Editor will check whether the submitted article meets the requirements and editorial policy of the journal [5], these activities will keep quality of the journal in supply new knowledge to the world science. Scholarly or scientific papers usually have certain pieces of metadata (usually assigned by authors) describing the topics and the main ideas of the contents [6].

The term *metadata* has been increasingly adopted and co-opted by more diverse audiences, the definition of what constitutes metadata has grown in scope to include almost anything that describes anything else [7]. Metadata are literally or technically 'data about data' or information about information or information that makes data useful. More over metadata as data whose primary purpose is to describe, define and/or annotate other data that accompanies it [8]. The structured data of Metadata describes the characteristics of a resource. It shares many similar characteristics to the cataloguing that takes place in libraries, museums and archives. The term "meta" derives from the Greek word denoting a nature of a higher order or more fundamental kind. A metadata record consists of a number of pre-defined elements representing specific attributes of a resource, and each element can have one or more values [9]. It is an extensive and expanding subject that is prevalent in many environments [10]. They provide information on such aspects as the 'who, what, where and when' of data and can be considered from the perspective of both the data producer and the data consumer. For the producer, metadata are used to document data in order to inform prospective users of their characteristics, while for the consumer, metadata are used to both discover data and assess their appropriateness for particular needs – their so-called 'fitness for purpose'. Providing metadata is the responsibility of each data provider with the quality of the metadata a significant problem [11]. Figure 1 shows some of metadata schemes.

In term of search, metadata is very useful key for search engine to recognize as the guide about what information should be provided to the users and it also determines the level of success of a search.

The most efficient way to make search work better is to bring some metadata to bear on the problem [12], because metadata are used for searching [13], and scientific papers usually have certain pieces of metadata (usually assigned by authors) describing the topics and the main ideas of the contents [6], and for full-text search, topic metadata is the right solution [14].

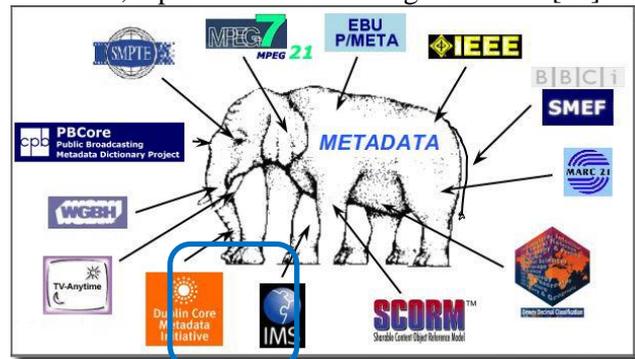

**Figure 1.** Metadata schemes [1]

Among some metadata standards, Dublin Core (DC) Metadata standard is one of the solid efforts [15]. Dublin Core is an international and interdisciplinary metadata standard that has been adopted by an array of communities wanting to facilitate resource discovery and build an interoperable information environment [16]. DC consists of three groups and 15 elements: 1) Content (title, subject and keywords, description, source, language, relation, coverage), 2) IP (creator, publisher, contributor, rights), and 3) Particular instance (date, type, format, identifying) [13]. The 15 elements of DC metadata are as follows: "dc:title"; (2) "dc:creator"; (3) "dc:subject"; (4) "dc:description"; (5) "dc:publisher"; (6) "dc:contributor"; (7) "dc:date"; (8) "dc:type"; (9) "dc:format"; (10) "dc:identifier"; (11) "dc:source"; (12) "dc:language"; (13) "dc:relation"; (14) "dc:coverage"; and (15) "dc:rights". Right now there are several additional elements of DC: Audience, Provenance, RightsHolder, InstructionalMethod, AccrualMethod, AccrualPeriodicity, AccrualPolicy [17]. Several reasons in using the DC are 1) The Dublin Core is a 15-element metadata element set proposed to facilitate fast and accurate information retrieval on the Internet [15], 2) DC will be more widely implemented in the future because of huge support from many international institutions such as Online Computer Library Centre (OCLC) and The Library of Congress [18], 3) Further, DC schema is flexible, easily understood and can be used to represent a variety of resources [19]. Another popular metadata scheme is IEEE LOM.

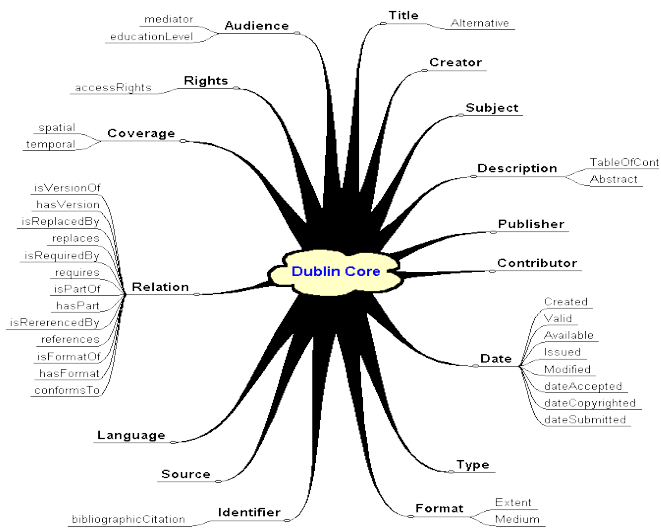

**Figure 2.** Dublin-Core metadata standard [2]

Portable Document Format (PDF[3]) is the global standard for capturing and reviewing rich information from almost any application on any computer system and sharing it with virtually anyone, anywhere. In recently publication, PDF documents become the standard *de-facto* for documents in digital libraries [20]. One possibility to identify a PDF file is extracting the title directly from the PDF's metadata [21]. At the moment, Adobe enriches PDF with XMP (has been introduced with Adobe Acrobat 5.0 and PDF 1.4 in April 2001). Adobe's eXtensible Metadata Platform (XMP[4]) is a labeling technology that allows us to embed data about a file, known as metadata or PDF metadata, into the file itself. XMP metadata travels with the file, and can be embedded in many common file formats including PDF, TIFF, and JPEG [5]. The XMP specification includes several schemas, but the most widely used predefined XMP schema is Dublin Core (DC). With XMP, reading metadata in a file is always the same [22]. XMP keeps the embedded metadata consistent. The XMP will always folowing the PDF file. One, we have the the article in PDF, then we will get the metadata as well. We could imagine the XMP similar to the role of DNA in our body.

In this paper, author develop and discuss a tool to harvest metadata from scientific journals published in PDF. Author also provides extra harvested information that should be useful to enrich the information about a particular article.

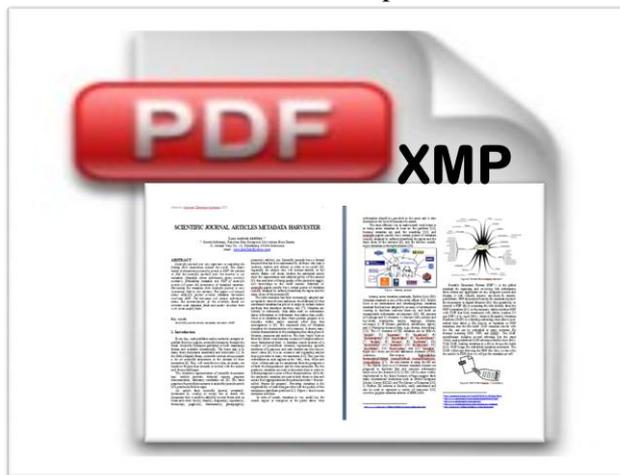

**Figure 3.** A journal article in PDF with XMP (Ilustration)

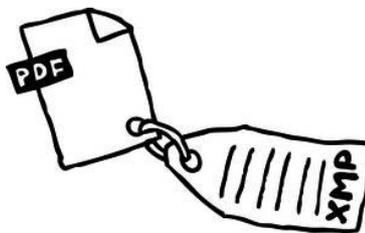

**Figure 4.** PDF XMP metadata [6] (Ilustration).

The rest of this paper will cover materials and method in Section 2, followed by results and discussion in Section 3, and conclude in Section 4.

**MATERIALS AND METHOD**

This paper will describe the scientific journal article metadata, XMP, harvester from PDF documents.

The experiments of this research need the collection of PDF documents from scholarly literatures. Author needs to collect those documents from scientific repository scholar repositories. Author uses personal collection of PDF articles about metadata which are downloaded from various scientific repositories or journals (ACM, IEEE, Springer, MATRIK, etc.). Those documents are published from 1998 until 2012.

In this work, the representation of a documents are full-text articles started with title, author(s), abstracts, keywords, metadata and will be ended by the list of reference.

In this paper, metadata are used to identify the information about the document related to author information, title, year of publicity, and PDF file information. Author adds some useful fields, such as; File name, File size, File page, File location, and Recency.

$$Recency = CurrentDate - CreationDate$$

Author develops this harvester by using popular Java programming language supported by ICEpdf. ICEpdf (by *IceSoft*) is an open source Java PDF engine that can render, convert, or extract PDF content within any Java application on a Web server [23]. Author develop the harvester based on this library and enrich with some useful fields.

## RESULTS AND DISCUSSION

Harvesting metadata is essential to get the hidden information from the PDF article, stored as XMP. Extracting and creating metadata for electronic documents help to arrange documents in a scientific way and support users can search them easily [17]. Some repositories are freely available for users to conduct some experiments, and every repository may provide different metadata in some format or types, such as: (1) RIS; (2) Plain Text; (3) Enw; or (4) BibTex. Those file formats are seperated from the PDF file. XMP are embedded in PDF. It means where ever we put the PDF file, then the XMP will exist with it.

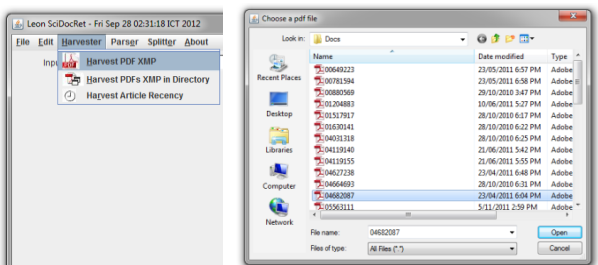

**Figure 5.** PDF XMP harvester.

To harvest the metadata from the repositories, we need tool to extract the information about the documents. In this paper, author develops a harvester to harvest metadata information from PDF article(s).

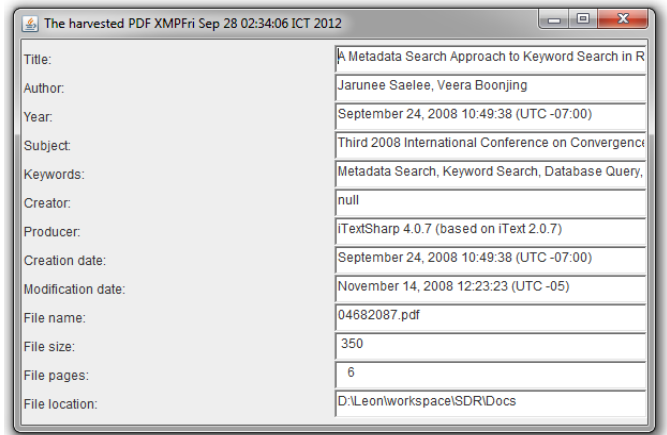

**Figure 6.** PDF XMP harvester per article

Figure 5-7 show the results of harvested PDF XMP from one article and many articles. The harvester able to retrieve all PDF XMP fields plus four additional fields that author add.

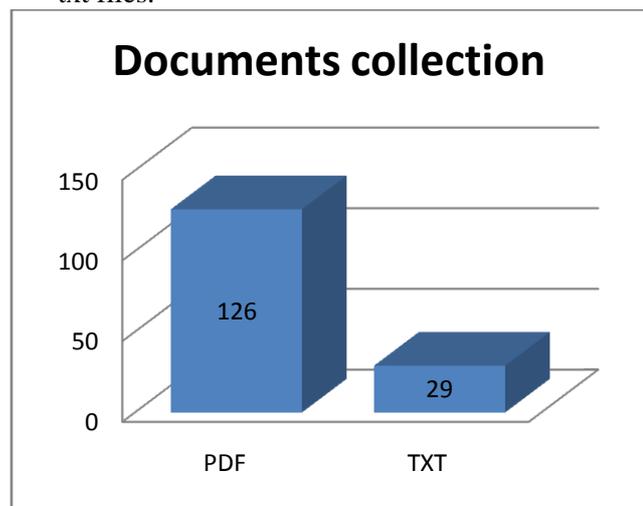

**Figure 7.** PDF XMP harvester in collections.

According to the experiment results, not all PDF documents are supplied with XMP metadata. The collection consist of 81.29% PDF and 18.71% txt files.

**Documents collection**

| | PDF | TXT |
|---|---|---|
| 126 | | 29 |

**Figure 8.** Percentage of documents files collection.

Among those PDF files, author focused on three main fields of PDF XMP, 1) year, 2) author, and 3) title, plus one additional fields of filename.

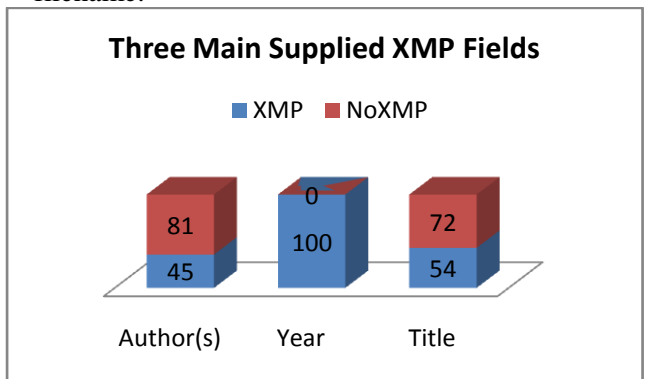

**Figure 9.** Three main PDF XMP fields

Author use these three fields because these three fields very important for researchers to recognize the scientific journal articles.

Based on all PDF files in the collections, we can analyze: 1) 45% of the articles are supplied with the author field, 2) 42.9% of the articles are supplied with the the title field, and 3) 100% of articles have their year field. The percentage of recency field is equal to year field (100%), because recency formula is CurrentDate − CreationDate. And last but not least, the additional fields, filename and recency, are 100% harvested, because these fields are added by the author of this harvester.

**Table 1.** The percentage of PDF XMP fields supplied by it's author(s)/publisher

| Fields | Percent (%) | Note |
|---|---|---|
| Filename | 100 | Additional field |
| Year | 100 | XMP field |
| Recency | 100 | Additional field |
| Author | 45 | XMP field |
| Title | 42.9 | XMP field |

## CONCUSION

Metadata are very useful to enrich the scientific journal article. Some elements of scientific journal such as author, title, and year. Metadata could stored in several file formats, such as; RIS; (2) Plain Text; (3) Enw; or (4) BibTex. Another scheme to store the metadata is using

XMP technology when the article is in PDF format. These information will be embedded in PDF article as hidden information or document properties. These hiden information consist of valuables information that summarize the contents of article. PDF format become standard for disseminate scientific finding.

- This harvester able to retrieve all of XMP fields from PDF files
- Author enriches this harvester with some useful additional fields beside XMP, such as recency
- The added recency field could be used to count the age of an article
- XMP technology of PDF become new standard to store the metadata information of ascientific article for the future
- At the moment not all articles published in PDF format are supplied by their author(s)/publisher with metadata in XMP. This is a challenge for next research.

**Leon Andretti Abdillah**, He earned bachelor degree in Computer Science, Study Program of Information Systems from STMIK Bina Darma in 2001, and Master in Management, Concentration of Information Systems from Universitas Bina Darma in 2006. He ever continue his PhD study at The University of Adelaide (2010-2012) in School of Computer Science. At the moment, he works as lecturer at Bina Darma University, in Information Systems study program. His main research interests are Information Systems, Scientific Journal, Information Retrieval, Human Resource IS, Database Systems, Programming, and Entrepreneur.